# Spin Hall effect associated with SU(2) gauge field

Yong Tao[†]

School of Economics and Business Administration, Chongqing University, Chongqing 400044, China

**Abstract.** In this paper, we focus on the connection between spin Hall effect and spin force. Here we investigate that the spin force due to spin-orbit coupling, which in two-dimensional system is equivalent to forces of Hirsch and Chudnovsky besides constant factors 3 and 3/2 respectively, is a part of classic Anandan force, and that the spin Hall effect is an anomalous Hall effect. Furthermore, we develop the method of AC phase to derive the formula for the spin force, and find that the most basic spin Hall effect originates from the AC phase and is therefore an intrinsic quantum mechanical property of spin. This method differs from approach of Berry phase in the study of anomalous Hall effect, which is the intrinsic property of the perfect crystal. On the other hand, we use an elegant skill to show that the Chudnovsky-Drude model is reasonable, and further have improved the theoretical values of spin Hall conductivity of Chudnovsky. Compared to the theoretical values of spin Hall conductivity in the Chudnovsky-Drude model, ours are in better agreement with experimentation. Finally, we discuss the relation between spin Hall effect and fractional statistics.

**Keyword.** Spin Hall effect; Gauge field; Fractional statistics; Aharonov-Casher effect.

**PACS number.** 72.25.-b; 71.70.Ej; 73.43.-f; 11.15.-q

## 1. Introduction

Based on the potential application of spin manipulation to the semiconductor industry, spintronics is rapidly developing and has been an important field of condensed matter physics [1]. Recently, more attention has been focused on the intrinsic spin Hall effect [2] related to spin-orbit coupling, and many mechanism have been proposed in order to explain the spin Hall effect [3-5]. Most importantly, the spin current has been observed by experiment [6]. Since Aharonov et al [7] proposed the Aharonov-Casher phase (AC phase) and Anandan et al [8a,9] realized that the spin-orbit coupling could be regarded as $SU(2)$ gauge field, the spin Hall effect by regarding spin as a gauge charge, in analogy with the charge Hall effect, had been studied within field theory and topology [10]. However, spin Hall effect was only regarded as a theoretical assumption before Hirsch pointed out a feasible scheme [11a,11b]. By studying the anomalous Hall effect, Hirsch noticed that there should be a net force on a moving spin in a periodic lattice, and this force would induce the spin Hall effect. Hirsch's work has attracted a lot of attention and promoted the study of spin Hall effect into a springtide. In order to obtain a universal theory to explain the spin Hall effect, Chudnovsky [3a] proposed a simple and insightful mechanism involving an extension of Drude model. Theoretical values of spin Hall conductivity computed within Chudnovsky-Drude model (C-D model) are in agreement with experimental data. Furthermore, the C-D model led to some interesting work [12]. More recently, however, the C-D model has been doubted. Kravchenko [13a] argued that the Eq. (24) in Ref. [3a] should be $\nabla^2 \Phi_0 = 0$ due to the electrical neutrality of conductors, where $\Phi_0$ is the total electrostatic potential; therefore, spin Hall conductivity in [3a] should be zero. Chudnovsky has responded [3b], but they still cannot come to an agreement [13b]. In this paper we will try to clarify the dilemma pointed out by Kravchenko [13a] because of the importance of C-D model.

---

[†] Corresponding author.
 E-mail address: taoyingyong@yahoo.com

Chudnovsky constructs the spin-dependent force by using the spin-orbit coupling, which implies that the C-D model may be related to the Goldhaber-Anandan gauge theory [8a,9]. This fact reminds us whether or not the origin of spin Hall effect is due to AC phase? Here we may as well guess, in analogy with the fact that the charge Hall effect originates from the AB phase [14], that the spin Hall effect originates from the AC phase and is therefore an intrinsic quantum mechanical property of spin. To see this, we can use an example to demonstrate the connection between spin Hall effect and AC phase. For example, He [15a,15b] has pointed out that one of the conditions inducing topological AC phase in the wave function is that there are regions where $\nabla \cdot \mathbf{E} \neq 0$, where $\mathbf{E}$ is electric field strength which can be denoted by $\nabla \Phi_0$. This is obviously different from the AB phase which is due to $\nabla \times \mathbf{A} \neq 0$, where $\mathbf{A}$ is potential vector. Accordingly if there is $\nabla \cdot \mathbf{E} = 0$ for any regions, then by Ref. [15a] the AC phase will be zero. Not only so, by the conclusion of Krachenko [13a,13b] the spin Hall effect will vanish as well. Although this example implies that there may be some relation between spin Hall effect and AC phase, it is hard to believe zero spin Hall effect since the spin Hall effect, in the metal [6], has been observed by experiment!

This paper has three main goals. First, we present a formula for spin force in two-dimensional system. This formula will give the theoretical value for the spin Hall conductivity. Second, we confirm that the most basic spin Hall effect indeed originates from AC phase and is therefore an intrinsic quantum mechanical property of spin. Third, we use an elegant skill to show that the C-D model is reasonable.

The paper is organized as follows. In section 2 we present a formula for spin force in two-dimensional system by using the Goldhaber-Anandan gauge theory for low energy electron. This formula gives the theoretical value for the spin Hall conductivity. In section 3 we develop the method of AC phase to derive the formula for the spin force. The method, which is distinguished from the approach of Berry phase, confirms that the most basic spin Hall effect originate from the AC phase and is therefore an intrinsic quantum mechanical property of spin. In section 4 we prove that the expression $\nabla \cdot \mathbf{E} = 0$ is not the necessary condition of electrical neutrality. In section 5 we use an elegant skill to confirm $\nabla \cdot \mathbf{E} = 4\pi\rho$ in the expression for the spin force. This skill is independent of the condition of electrical neutrality. Next, we discuss the relation between spin Hall effect and fractional statistics. Finally, section 6 presents conclusions.

## 2. The spin force in the Goldhaber-Anandan gauge framework

In accordance with the Ref. [10] we discuss the charge and spin Hall effect in a unitary framework, namely, the Goldhaber-Anandan gauge theory.

## 2.1 The Goldhaber-Anandan gauge theory

By using the Foldy-Wouthuysen transformation [16], the Hamiltonian in the low energy weak field limit of Dirac equation coupled to electromagnetic field can be written, i.e., the Eq. (25) in [8a], in the form:

$$H = \frac{1}{2m}\left(P - \frac{e}{c}\mathbf{A} - \frac{1}{2c}\boldsymbol{\mu}\times\mathbf{E}\right)^2 + eA_0 - \boldsymbol{\mu}\cdot\mathbf{B}.$$

Where $\boldsymbol{\mu} = \mu\cdot\boldsymbol{\sigma}$ is magnetic moment vector and $\boldsymbol{\sigma} = (\sigma^{(k)})$, $(k=1,2,3)$, are Pauli matrix.

Hence, the electromagnetic interaction is like as if it is due to an $SU(2)\otimes U(1)$ gauge field. Here the interaction of the spin in the electromagnetic field behaves as if the spin is a gauge charge and the interaction is due to the $SU(2)$ gauge fields. However, the $SU(2)$ gauge fields $\left(\boldsymbol{\mu}\cdot\mathbf{B}\quad \frac{\boldsymbol{\mu}}{2}\times\mathbf{E}\right)$ are not new physical variables but are induced by electromagnetic field. According to Anandan's idea [8a] we can construct a convenient Lagrangian which owns $SU(2)\otimes U(1)$ symmetry as follow:

$$L = \frac{i\hbar}{2}\left[\psi^+(D_0\psi) - (D_0\psi)^+\psi\right] - \frac{\hbar^2}{2m}(D_i\psi)^+(D_i\psi) - \frac{1}{16\pi}F_{\mu\nu}F^{\mu\nu}, \tag{1}$$

where $D_0 = \frac{\partial}{\partial x^0} + i\frac{\mu}{\hbar}W_0^{(k)}\sigma^{(k)} + ieA_0$, $D_i = \frac{\partial}{\partial x^i} + i\frac{\mu}{\hbar c}W_i^{(k)}\sigma^{(k)} + ieA_i$,

$$W_\mu^{(1)} = \left(-B_x \quad 0 \quad -\frac{E_z}{2} \quad \frac{E_y}{2}\right), \quad W_\mu^{(2)} = \left(-B_y \quad \frac{E_z}{2} \quad 0 \quad -\frac{E_x}{2}\right),$$

$$W_\mu^{(3)} = \left(-B_z \quad -\frac{E_y}{2} \quad \frac{E_x}{2} \quad 0\right), \quad F_{\mu\nu} = \partial_\mu A_\nu - \partial_\nu A_\mu,$$

and $\psi = \begin{pmatrix}\psi_\uparrow \\ \psi_\downarrow\end{pmatrix}$ denotes two-components spinor.

It is easily shown that there is no purely $SU(2)$ gauge term in the Lagrangian (1). However, we can still imitate Yang-Mills field [17] to construct the $SU(2)$ spin gauge field strength $G_{\mu\nu}^{(i)} = \partial_\mu W_\nu^{(i)} - \partial_\nu W_\mu^{(i)} + \frac{2\mu}{\hbar c}f_{ijk}W_\mu^{(j)}W_\nu^{(k)}$. It is worth noticing that, recently, Jin et al [18] had supposed there exist purely $SU(2)$ gauge term and hence constructed a complete $SU(2) \otimes U(1)$ gauge model. The advantage of their work is that the non-conservation of spin current in the presence of Yang-Mills field is due to its non-Abelian feature. Stated another way, the spin gauge field will take a part of spin charge. If we construct total current [17] $\vartheta_\mu^{(i)} = J_\mu^{(i)} + \frac{2\mu}{\hbar c} f_{ijk} W_\nu^{(j)} G_{\mu\nu}^{(k)}$, then the conserved relation holds, i.e., $\partial_\mu \vartheta_\mu^{(i)} = 0$. However, the spin current is non-conserved, i.e., $\partial_\mu J_\mu^{(i)} \neq 0$.

## 2.2 The spin force for two-dimensional system

As is well known, the charge Hall effect is due to the Lorentz force that acts on a moving charge. Likewise, Hirsch [11a] pointed out that the force on a moving spin in the electric field plays an important role in studying the connection between anomalous Hall effect and spin Hall effect. Later, Shen [19] and Chudnovsky [3a] also realized the importance of spin-dependent force in the spin Hall effect. Chudnovsky made an important step in this regard by constructing the C-D model involving spin-dependent force. The theoretical values of spin Hall conductivity computed within the C-D model are in agreement with the experimental data. On the other hand, according to the Goldhaber-Anandan gauge theory, the spin can be regarded as a gauge charge due to $SU(2)$ spin gauge field. This shows that the spin should be exerted by classical Anandan force $F_\mu$ in terms of $SU(2)$ spin gauge field [8a,8b], where $F_\mu = G_{\mu\nu}^{(i)} J_\nu^{(i)}$.

Next we will find that the Anandan force in the two-dimensional system is equivalent to the spin force of Chudnovsky, provided that the symmetry condition[1] $\left\langle \frac{\partial^2 \Phi_0}{\partial x^i \partial x^j} \right\rangle = A \delta_{ij}$ and the term of Pauli interaction are dropped. To this end, there have two steps. First, we observe the Lorentz force $f_y$ in Fig. 1.

---

[1] The symmetry condition is given by Eq. (15) in [3a], where $\Phi_0$ is the total electrostatic potential and $A$ is a constant. This symmetry condition indicates, by observing electric field strength $\mathbf{E} = \nabla \Phi_0$, that

$\frac{\partial E_x}{\partial x} = \frac{\partial E_y}{\partial y} = \frac{\partial E_z}{\partial z} = A$.

Adding the magnetic field $B_z$ project along the direction $z$, and the electric field $E_x$ along the direction $(-x)$.

The electron move with velocity $v_x$ and charge $(-e)$ project along the direction $x$.

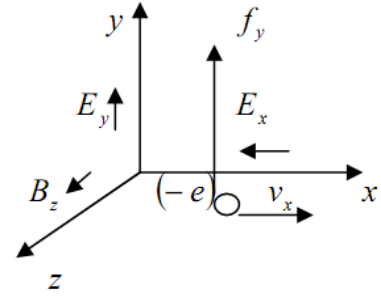

FIG.1

The charge Hall effect is determined by the requirement that the Lorentz force $f_y = 0$ in the direction $y$. However, there would be spin force, which is perpendicular to the electric field $E_x$, from the spin-orbit interaction [20]. Only the composition of Lorentz force and spin force is zero, there would be new equilibrium in the direction $y$. Second, we consider the Anandan force $F_y$ in Fig. 2.

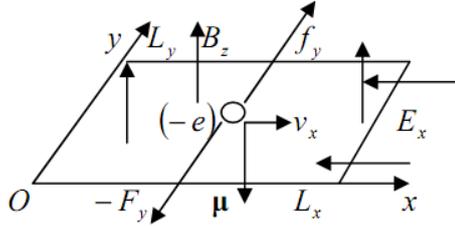

FIG.2

The width and length of the two dimensional $x-y$ plane are $L_y$ and $L_x$ respectively. The electron with velocity $v_x$ and charge $(-e)$ is moving in the $x-y$ plane, in the $y$ direction it will be acted by the Lorentz force $f_y$ and the Anandan force $(-F_y)$. The direction of magnetic moment vector $\mu$ of electron is anti-parallel to $B_z$.

The physical picture for the two-dimensional electron system involving the Anandan force is illustrated in Fig 2. Later we will find that the Anandan force has a particularly simple form. To this end, we start to calculate the Anandan force $F_y$ in the direction $y$, where $F_y = F_2 = G^{(i)}_{2\nu} J^{(i)}_\nu$. It is easy to compute the relations

$$G^{(1)}_{2\nu} J^{(1)}_\nu = \left[ \frac{\partial B_x}{\partial y} - \frac{1}{2}\frac{\partial E_z}{\partial t} - \frac{\mu}{\hbar c} B_y E_x \right] J^{(1)}_0 + \left[ -\frac{1}{2}\frac{\partial E_z}{\partial x} + \frac{\mu}{2\hbar c} E_x E_z \right] J^{(1)}_1$$
$$+ \left[ -\frac{1}{2}\left(\frac{\partial E_y}{\partial y} + \frac{\partial E_z}{\partial z}\right) - \frac{\mu}{2\hbar c} E_x^2 \right] J^{(1)}_3$$

,

$$G_{2\nu}^{(2)} J_\nu^{(2)} = \left[ \frac{\partial B_y}{\partial y} + \frac{\mu}{\hbar c} (B_z E_z + B_x E_x) \right] J_0^{(2)} + \left[ -\frac{1}{2} \frac{\partial E_z}{\partial y} + \frac{\mu}{2\hbar c} E_y E_z \right] J_1^{(2)}$$
$$+ \left[ \frac{1}{2} \frac{\partial E_x}{\partial y} - \frac{\mu}{2\hbar c} E_x E_y \right] J_3^{(2)},$$

$$G_{2\nu}^{(3)} J_\nu^{(3)} = \left[ \frac{\partial B_z}{\partial y} + \frac{1}{2} \frac{\partial E_x}{\partial t} - \frac{\mu}{\hbar c} B_y E_z \right] J_0^{(3)} + \left[ \frac{1}{2} \left( \frac{\partial E_x}{\partial y} + \frac{\partial E_y}{\partial y} \right) + \frac{\mu}{2\hbar c} E_z^2 \right] J_1^{(3)}$$
$$+ \left[ \frac{1}{2} \frac{\partial E_x}{\partial z} - \frac{\mu}{2\hbar c} E_x E_z \right] J_3^{(3)}.$$

Here, we observe that $E_z = B_x = B_y = 0$ and $J_3^{(i)} = 0$ in this two-dimensional system. Therefore, the Anandan force in the direction $y$ can be derived as

$$F_y = \frac{\partial B_z}{\partial y} J_0^{(3)} + \frac{1}{2} \left( \frac{\partial E_x}{\partial x} + \frac{\partial E_y}{\partial y} \right) J_1^{(3)}. \tag{2}$$

The first term $\frac{\partial B_z}{\partial y} J_0^{(3)}$ in Eq. (2) denotes Pauli interaction, which has been confirmed by Stern-Gerlach experiment. In this paper out attention will be concentrated on the second term, which is attributed to the spin-orbit interaction and could give the spin force $F_y^{spin}$. If we use Chudnovsky's assumption [3a] $\nabla \cdot \mathbf{E} = 4\pi\rho$ and the expression for the spin current [1,8a,11a] $J_1^{(3)} = -\frac{\mu}{c}(n_\uparrow - n_\downarrow) v_x$, then the spin force $F_y^{spin}$, by observing that $E_z = 0$ in this two-dimensional system which leads to $\frac{\partial E_x}{\partial x} + \frac{\partial E_y}{\partial y} = \nabla \cdot \mathbf{E} = 4\pi\rho$, can be obtained from the second term in Eq. (2), where

$$F_y^{spin} = -\frac{2\pi\rho\mu}{c}(n_\uparrow - n_\downarrow) v_x. \tag{3}$$

It is carefully noticed that the direction of spin force, in the Fig 2, is in the direction $\mathbf{v} \times \boldsymbol{\mu}$ with $\mathbf{v}$ the velocity vector. The Eq. (3) is the main result of this paper. It can be also obtained from Eq. (8) in Ref. [3a] provided that we drop the symmetry condition $\left\langle \frac{\partial^2 \Phi_0}{\partial x^i \partial x^j} \right\rangle = A \delta_{ij}$. Conversely, if the symmetry condition, as Chudnovsky has done [3a], holds, then there would be, by the footnote 1, that $\frac{\partial E_x}{\partial x} + \frac{\partial E_y}{\partial y} = \frac{2}{3} \nabla \cdot \mathbf{E} = \frac{8}{3}\pi\rho$, which implies, by observing the second term in Eq. (2), that the spin force $F_y^{spin}$ is equivalent to the result of Chudnovsky multiplied by a factor of $\frac{3}{2}$. In addition, the Eq. (10) in [3a] also indicates, in accordance with the Eq. (25) in [8a], that the spin-orbit coupling could be regarded as $SU(2)$ spin gauge field. Hence, the sign of our spin force $F_y^{spin}$ agrees with the result of Chudnovsky. Recently [11c], Hirsch realized that the spin force of Chudnovsky was equivalent to the force that he had ever obtained in Ref. [11a]. Chudnovsky [3a] also pointed out that his spin force only coincided up to a factor of 2 with the result obtained by Hirsch. Therefore, our spin force $F_y^{spin}$ is as well equivalent to results of Hirsch and Chudnovsky besides constant factors 3 and $\frac{3}{2}$ respectively. However, we must point out that our result is rigorous for two-dimensional system.

## 2.3 The spin Hall conductivity

In this subsection, we use the expression for the spin force, i.e., Eq. (3), to derive the spin Hall conductivity. It is shown in Fig. 2 that the equilibrium of force along the direction $y$ determines Hall conductivity. In the direction $y$ the Lorentz force reads $f_y = \rho\left(E_y - \frac{v_x}{c}B_z\right)$. Note that $F_y^{spin} + f_y = 0$ and Eq. (3), one obtains $v_x = \frac{cE_y}{B_z + 2\pi M}$, where the magnetization $M = (n_\uparrow - n_\downarrow)\mu$.

Therefore, the Hall conductivity can be obtained as

$$\sigma_{xy} = \frac{nev_x}{E_y} = \frac{nec}{B_z + 2\pi M}. \qquad (4)$$

The corresponding resistivity reads $\rho_{xy} = \frac{1}{nec}B_z + 4\pi\frac{1}{2nec}M$, which agrees with the empirical formula [11a,11b] $\rho_H = R_0 B_z + 4\pi R_s M$. The fact shows that the spin Hall effect is an anomalous Hall effect [2].

If one introduces $\frac{2\pi M}{B_z} < 1$, Eq. (4) can be written as

$$\sigma_{xy} = \frac{nec}{B_z}\left[1 - \frac{2\pi M}{B_z} + \left(\frac{2\pi M}{B_z}\right)^2 - ...\right]$$

In the spirit of Chudnovsky [3a], we can regard the spin Hall conductivity $\sigma_{xy}^s$ as two order perturbation of the Hall conductivity. Thus, the charge Hall conductivity reads $\sigma_{xy}^c = \frac{nec}{B_z}$, and the corresponding spin Hall conductivity reads $\sigma_{xy}^{s} = \dfrac{nec}{B_z}\left[-\dfrac{2\pi M}{B_z} + \left(\dfrac{2\pi M}{B_z}\right)^2 - ...\right]$. Note that $n = n_\uparrow + n_\downarrow$, substituting $\sigma_{xy}^{c}$ into $\sigma_{xy}^{s}$ yield:

$$\sigma_{xy}^{s} \approx -\dfrac{n_\uparrow - n_\downarrow}{n_\uparrow + n_\downarrow} \times \dfrac{\pi\hbar}{mc^2}\left(\sigma_{xy}^{c}\right)^2, \tag{5}$$

where we keep the first term in $\sigma_{xy}^{s}$ only.

If the carriers, as shown in Fig. 2, are all polarized in the same direction, then Eq. (5) reads

$$\sigma_{xy}^{s} \approx \dfrac{\pi\hbar}{mc^2}\left(\sigma_{xy}^{c}\right)^2. \tag{6}$$

Eq. (6) coincides up to a factor of $\dfrac{3}{2}$ with the result obtained by Chudnovsky [3a], since we have dropped the symmtry condition $\left\langle\dfrac{\partial^2 \Phi_0}{\partial x^i \partial x^j}\right\rangle = A\delta_{ij}$ as previous discussion in subsection 2.2.

TABLE1. Comparison of Eq. (6) with experimental data [6] on ordinary and spin Hall conductivity in Al strips of 12 and 25 nm thickness.

| Conductivity | Experiment $(\Omega^{-1}m^{-1})$ | Theory $(\Omega^{-1}m^{-1})$ |
| --- | --- | --- |
| $\sigma_{xy}^{c}$ (12nm) | $1.05 \times 10^7$ | |
| $\sigma_{xy}^{s}$ (12nm) | $(3.4 \pm 0.6) \times 10^3$ | $3.9 \times 10^3$ |
| $\sigma_{xy}^{c}$ (25nm) | $1.7 \times 10^7$ | |
| $\sigma_{xy}^{s}$ (25nm) | $(2.7 \pm 0.6) \times 10^3$ | $10.35 \times 10^3$ |

From Table.1 we notice that, the thinner material, the better theoretical values conform to experimental data. For this situation, we explain that our results only are accuracy in the purely two-dimensional system. Unfortunately, in the three-dimensional system, by observing $E_z \neq 0$ from the Anandan force $F_z \neq 0$ along the direction $z$, one obtains $\dfrac{\partial E_x}{\partial x} + \dfrac{\partial E_y}{\partial y} \neq 4\pi\rho$.

This implies that we cannot have Eq. (3).

## 3. The spin Hall effect and AC phase

In this section we investigate the relation between spin Hall effect and AC phase, and confirm that the most basic spin Hall effect indeed originate from the AC phase and is therefore an intrinsic quantum mechanical property of spin. Here, for simplicity, we focus on the model of mesoscopic ring [20] (see Fig. 3 and 4).

In analogy with the fact that the time-dependent AB phase could induce the persistent electric current, Refs. [21,22] pointed out that the time-dependent AC phase would induce the persistent spin current. The latter could be called as the spin Faraday law. Next, the further step, we will point out that the AC flux coupled to the spin current, in analogy with the fact that the AB flux coupled to the electric current induces the Lorentz force, could induce the spin force. More importantly, we will find that the spin force may be regarded as the localized effect of AC phase.

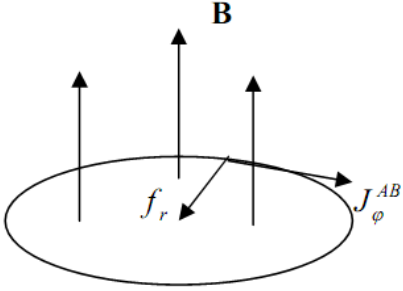
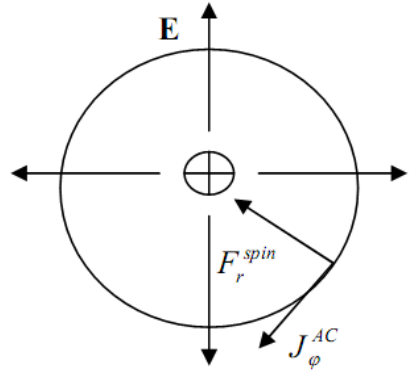

FIG.3　　　　　　　　　　　　　　　FIG.4

The relation between spin (electric) current and AC (AB) flux is given in [20]

$$j_\varphi^{AC} = -\frac{c}{4\pi R} Tr \frac{\partial H(\Phi_{AB} + \sigma^{(3)}\Phi_{AC})}{\partial \Phi_{AC}} \sigma^{(3)}, \tag{7}$$

$$j_\varphi^{AB} = -\frac{c}{2\pi R} \frac{\partial H(\Phi_{AB} + \sigma^{(3)}\Phi_{AC})}{\partial \Phi_{AB}}, \tag{8}$$

with spin current $j_\varphi^{AC}$ and electric current $j_\varphi^{AB}$ given by

$j_\varphi^{AC} = \mu(n_\uparrow - n_\downarrow)v_\varphi$ and $j_\varphi^{AB} = e(n_\uparrow + n_\downarrow)v_\varphi$ respectively.

Here $H = H(\Phi_{AB} + \sigma^{(3)}\Phi_{AC})$ is energy, $\Phi_{AB} = -\oint \mathbf{A} \cdot \mathbf{dl}$ is AB flux, $\Phi_{AC} = \frac{1}{2}\oint \boldsymbol{\sigma} \times \mathbf{E} \cdot \mathbf{dl}$ is AC flux, $v_\varphi$ is the velocity along the ring, and $R$ is the radius of ring.

The expression for the spin force can be derived from Eq. (7). To this end, we rewrite Eq. (7) as

$$j_\varphi^{AC} = -\frac{c}{4\pi R} Tr \frac{\partial H}{\partial S} \frac{\partial S}{\partial \Phi_{AC}} \sigma^{(3)}, \tag{9}$$

where $S$ is the area of ring.

Note that $\oint dl = 2\pi R$ and that Eq. (9) is equivalent to $\frac{\partial H}{\partial S} = -\frac{\partial \Phi_{AC}}{\partial S} \times \frac{2\pi R}{c} j_\varphi^{AC}$, one can obtain

$$\frac{\partial H}{\partial S} = -\frac{1}{c} \oint \frac{\partial (\Phi_{AC} j_\varphi^{AC})}{\partial S} dl. \tag{10}$$

Exert the differential operator $\nabla$ on Eq. (10) and let $H = \iint \boldsymbol{\varepsilon} \cdot \mathbf{dS}$, one obtains

$$F_r^{spin} = (\nabla \cdot \boldsymbol{\varepsilon})_r = -\frac{1}{c} \frac{\partial (\Phi_{AC} j_\varphi^{AC})}{\partial S}, \tag{11}$$

where $F_r^{spin}$ denotes the density of radial spin force and $\mathbf{dS}$ which is second-order tensor, denotes infinitesimal element of area.

It is carefully noticed that, in the Eq. (11), $\boldsymbol{\varepsilon}$ is the density of energy which is second-order tensor, and that $\nabla \cdot \boldsymbol{\varepsilon}$ is first-order tensor, where symbol "dot" denotes transvection.

Note that the AC phase factor reads $e^{i\frac{\mu}{\hbar c}\Phi_{AC}}$, we realize that the spin force $F_r^{spin}$, in the Eq. (11), is closely related to the AC phase, and that the AC flux coupled to the spin current does induce the spin force. It indicates that the topological origin of spin Hall effect is due to the AC phase.

Substitution of the expression for AC flux [15a] $\Phi_{AC} = \frac{1}{2} \iint \nabla \cdot E dS$ and the assumption of Chudnovsky $\nabla \cdot E = 4\pi\rho$ into Eq. (11) leads to

$$F_r^{spin} = -\frac{2\pi\rho\mu}{c}(n_\uparrow - n_\downarrow)v_\varphi. \tag{12}$$

Obviously, Eq. (12) is in agreement with Eq. (3).

Analogously, Eq. (8) gives the Lorentz force $f_r = -\frac{\rho}{c} B_z v_\varphi$.

From the discussion in subsection 2.3 we have known that Eq. (12) or Eq. (3) gives the spin Hall conductivity. Therefore, we have confirmed that the spin Hall effect indeed originates from the AC phase, which is purely intrinsic effect of spin. Here we must point out that our approach due to the AC (AB) phase is different from the approach of Berry phase [23a,23b], even though the AC and AB phases are special Berry phases [15b,24].

In fact, the approach of Berry phase, in Ref. [23a,23b], which is dependent on the band structure of the perfect crystal and is therefore an intrinsic property of the material, would perform more physical insight for intrinsic anomalous Hall effect as compared to that of the approach of AC phase. From the discussion of this section, however, we realize that the AC phase should be regarded as the most basic origin of the spin Hall effect since it is the intrinsic property of the spin. Therefore, the spin Hall effect due to the AC phase, as the most basic effect, should be distinguished from the skew scattering, side-jump and the Berry phase originating from the band structure.

## 4. The necessary condition of electrical neutrality

Although our theoretical values (Table. 1) are in accordance with experimental data, the assumption $\nabla \cdot \mathbf{E} = 4\pi\rho$, according to Ref [13a], is contrary to the electrical neutrality of conductor. However, next, we will prove that the expression $\nabla \cdot \mathbf{E} = 0$ is not the necessary condition of electrical neutrality. To the knowledge of author, the electron may be regarded as a geometric point due to the quantization of charge. However, this would induce that the electric field strength $\mathbf{E}$ has singularity in the conductor $\Omega$, so that the formula $\nabla \cdot \mathbf{E} = 0$ is invalid. If $\mathbf{E}$ has $n$ singular points in $\Omega$, the Gauss formula reads

$$\oiint_{\partial\Omega} \mathbf{E} \cdot \mathbf{dS} = \sum_{i=1}^{n} \oiint_{\delta_i} \mathbf{E} \cdot \mathbf{dS} + \iiint_{\Omega / \bigcup_i \delta_i} \nabla \cdot \mathbf{E} dV ,$$

where $\delta_i$ denotes the $i$ th infinitesimal open set that has covered the $i$ th singular point, and the Lebesgue measure of union set $\bigcup_i \delta_i$ is $L\left(\bigcup_i \delta_i\right) \leq \sum_i L(\delta_i) = 0$.

In fact, the rigorously necessary condition of electrical neutrality should be $\oiint_{\partial\Omega} \mathbf{E} \cdot \mathbf{dS} = 0$. Here, we had to notice that $\nabla \cdot \mathbf{E}$ could be invalid in the area $\bigcup_i \delta_i \subset \Omega$. To remove the singularity of $\mathbf{E}$ in $\bigcup_i \delta_i$, we may bring the Dirac function $\delta(\mathbf{x})$ to satisfy

$$\nabla \cdot \mathbf{E} = \sum_{i=1}^{n} q_i \delta(\mathbf{x} - \mathbf{x_i}) = 0, \text{ where } |q_i| \text{ is unit charge. However, the method cannot remove the}$$

singularity of $\mathbf{E}$ in $\Omega$, e.g. the electric dipole[2] in Fig 5. The electric field strength of electric dipole reads

$$\mathbf{E} = \lim_{|\mathbf{r_1}-\mathbf{r_2}|\to 0}[\mathbf{E}(\mathbf{r_1}) + \mathbf{E}(\mathbf{r_2})] = \lim_{|\mathbf{r_1}-\mathbf{r_2}|\to 0} \mathbf{E}(\mathbf{r_1} - \mathbf{r_2}) = \frac{3(\mathbf{p}\cdot\mathbf{r})\mathbf{r} - \mathbf{p}r^2}{4\pi r^5} .$$

Obviously, it not only satisfies the condition of electrical neutrality $\oiint_{\partial\Omega} \mathbf{E} \cdot \mathbf{dS} = 0$, but also gives non-zero charge density $\rho(\mathbf{r}) = \nabla \cdot \mathbf{E} = -(\mathbf{p} \cdot \nabla)\delta(\mathbf{r})$. Furthermore, this electrically neutral system can be exerted by an electric field force[3]:

---

[2] Similar to the structure of conductor, though the distance between positive and negative charge of electric dipole may approach zero, they are not annihilated

[3] Likewise, if we regard magnetic moment of neutron as localized current, it would be exerted by the same force [25]. Furthermore, the effect can induce the $SU(2)$ spin gauge field [26].

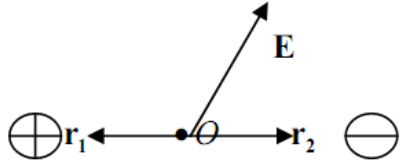

FIG.5

$$\mathbf{F}(0) = -\int (\mathbf{p}\cdot\nabla)\delta(\mathbf{r})\mathbf{E}(\mathbf{r})dV = -(\mathbf{p}\cdot\nabla)\mathbf{E}(0).$$

All these difficulties originate from the fact that we cannot find any area $\delta \subset \Omega$ in which $\nabla\cdot\mathbf{E}(\mathbf{r_1}-\mathbf{r_2})$ is uniform convergence in terms of $|\mathbf{r_1}-\mathbf{r_2}|$, so that [27]

$$\nabla\cdot\left(\lim_{|\mathbf{r_1}-\mathbf{r_2}|\to 0}\mathbf{E}(\mathbf{r_1}-\mathbf{r_2})\right) \neq \lim_{|\mathbf{r_1}-\mathbf{r_2}|\to 0}(\nabla\cdot\mathbf{E}(\mathbf{r_1}-\mathbf{r_2})).$$ More importantly, the singularity of field has universality in the case of approximation of single particle. Under the Born-Oppenheimer approximation, the separating of electron and nucleon wave functions will induce singular gauge field [28], even in the case for separating the degree of freedoms of spin and space [29]. Therefore, the expression $\nabla\cdot\mathbf{E} = 0$ is not the necessary condition of electrical neutrality. This implies that the conclusion of Krachenko is invalid. In next section, we will give the magnitude of $\nabla\cdot\mathbf{E}$.

## 5. Nonzero spin force and fractional statistics

### 5.1 Nonzero spin force

To prove that the C-D model is reasonable, in this subsection, we will use an elegant skill (that is, differential mapping) to determine magnitude of $\nabla\cdot\mathbf{E}$ in the expression for the spin force (Eq. (3) or Eq. (12)). To this end, we observe that the AC effect leads to nontrivial phase shifts and to a topological interference effect of the wave function of a particle with spin. Furthermore, we believe that this topological effect is geometrically invariant under differential mapping. On the other hand, in subsection 2.1, we have known that the $SU(2)$ spin gauge field is induced by electromagnetic field. These facts lead us to guess whether or not the $SU(2)$ spin gauge field could be mapped as the $U(1)$ electromagnetic field?

In fact, the similar question has been discussed rigorously by Anandan [8a]. Recently, there are some relevant works [30-32] again. The $SU(2)$ spin gauge field is reduced (mapped) to the correctional term of electromagnetic field, which can be shown as a part of the Maxwell equation in media. This implies that the spin Hall effect induced by $SU(2)$ spin gauge field would be correctional term of ordinary Hall effect.

In the spirit of Anandan [8a], varing $A_\mu$ in Lagrangian (1) gives field equation for $F_{\mu\nu}$

$$\frac{1}{4\pi}\partial_\mu F^{\mu 0} = \varepsilon^{ijk}\partial_i J^{i(k)} + \rho_0^c, \tag{13}$$

$$\frac{1}{4\pi}\partial_\mu F^{\mu i} = -\varepsilon^{ijk}\frac{1}{c}\partial_0 J^{j(k)} - \varepsilon^{ijk}\partial_j J^{0(k)} + \frac{1}{c}\mathbf{j_0^c}, \tag{14}$$

with $J^{0(k)} = \mu\psi^+\sigma^{(k)}\psi$ and $J^{i(k)} = -\frac{i\hbar\mu}{4mc}\left[\psi^+\sigma^{(k)}(D_i\psi) - (D_i\psi)^+\sigma^{(k)}\psi\right]$,

where $J^{i(k)}$ denotes the spin current induced by the spin-orbit coupling. In the WKB approximation [8a],

$$J^{0(k)} \approx \mathbf{M} \quad \text{and} \quad J^{i(k)} \approx \frac{1}{2c}\mathbf{v}\otimes\mathbf{M}.$$

One can introduce $\mathbf{p} = \frac{1}{2c}\mathbf{v}\times\mathbf{M}$, where $\mathbf{p}$ is polarization and $\mathbf{M}$ is magnetization.

Then, by Eqs. (13) and (14), $\mathbf{H} = \mathbf{B} - 4\pi\times\mathbf{M}$ and $\mathbf{D} = \mathbf{E} + 4\pi\times\mathbf{p}$ would satisfy the Maxwell equations in media

$$\nabla\cdot\mathbf{D} = 4\pi\rho_0^c,$$

$$\nabla\times\mathbf{H} - \frac{1}{c}\frac{\partial\mathbf{D}}{\partial t} = \frac{4\pi}{c}\mathbf{j_0^c}.$$

Here we only focus on the electric vector

$$\mathbf{D} = \mathbf{E} + 4\pi\mathbf{p} = \mathbf{E} + \frac{2\pi}{c}\mathbf{v}\times\mathbf{M}. \tag{15}$$

The Eq. (15) shows that there exists additional electromagnetic force

$$\mathbf{f}' = \frac{2\pi\rho}{c}\mathbf{v}\times\mathbf{M}. \tag{16}$$

Obviously, the additional electromagnetic force (16), regardless of the magnitude or direction, is in agreement with the spin force in Eq. (3) or Eq. (12). This fact shows that the assumption of Chudnovsky [3a], which reads $\nabla\cdot\mathbf{E} = 4\pi\rho$, is right. In fact, there should be terms of nuclei wave functions in the Lagrangian (1) because of the structure of conductor. However, note that the mass of nuclei far exceed that of the electron, these terms have been removed by applying the Born-Oppenheimer approximation. According to the Born-Oppenheimer approximation, the dynamics of the nuclei may either be neglected, or treated classically as a slow, adiabatic motion, whereas the electron must be treated quantum mechanically. This is why there is no contribution of magnetic moment of nuclei in the Eq. (15). Because of this approximation, the magnetic moment of nuclei can be neglected, and the residual magnetic moment of electron results in an "effect" magnetic field. The fact is the key that we can get rid of the dilemma pointed by Kravchenko [13a]. If the magnetic moment of nuclei is equivalent to that of electron, there will be no additional electromagnetic force. However, this elegant skill cannot be applied in the C-D model in which the technique of differential mapping is neglected.

### 5.2 The fractional statistics

In previous subsection, Eq. (16) shows that $2\pi\mathbf{M}$ can be regarded as an 'effect' magnetic field. This point of view has also been implied in the discussion of Hirsch [11c]. Furthermore, next, we will point out that the 'effect' magnetic field can induce fractional statistics. Firstly, by the Fig. 2, one can introduce the Schrodinger equation involving 'effect' magnetic field

$$\frac{1}{2m}\left\{\left[\left(\hat{P}_x - \frac{2\pi e \mathrm{M}}{c} y\right)^2 + \hat{P}_y^2\right] + eEy\right\}\psi = \varepsilon\psi. \tag{17}$$

The solution of Eq. (17) can be derived as $\psi = \frac{1}{\sqrt{L_x}} \exp\left(\frac{i}{\hbar} P_x x\right)\chi(y - y_p),$ (18)

Where $y_p = \frac{c}{2\pi e \mathrm{M}}\left(P_x - \frac{Emc}{2\pi \mathrm{M}}\right).$

The corresponding Landau level reads

$$\varepsilon_k = \hbar\omega_c\left(k + \frac{1}{2}\right) + eEy_p + \frac{m}{2}\left(\frac{cE}{2\pi \mathrm{M}}\right)^2, \quad k = 0,1,2...$$

where the gyromagnetic ratio reads $\omega_c = \frac{2\pi e \mathrm{M}}{mc}$. By using Eq. (18) the average current density reads

$$j_x = \frac{\int \left(-\frac{P_x e}{m} + \frac{2\pi e^2 \mathrm{M} y}{mc}\right)|\chi(y - y_p)|^2 \frac{1}{L_x} dy}{L_y} = -\frac{necE}{2\pi \mathrm{M}}.$$

Thus, the corresponding spin Hall conductivity reads $\sigma_{xy}^s = \frac{j_x}{E} = -\frac{nec}{2\pi \mathrm{M}},$

which is in accordance with Eq. (4) as $B = 0$, i.e., $\sigma_{xy}(B = 0) = -\frac{nec}{2\pi \mathrm{M}}.$

Now, we bring a many-body wave function for the two-dimensional electron gases $\Psi = \Psi(r_1 s_1, r_2 s_2 ... r_N s_N, t)$, where $r_i$ and $s_i$ are the space and spin coordinate respectively.

Exerting an exchange operator $P_{ij}$ on $\Psi$, one, observing Eq. (17), obtains

$$P_{ij}\Psi(r_1 s_1, r_2 s_2 ... r_i s_i ... r_j s_j ... r_N s_N, t) = \exp(i\theta)\Psi(r_1 s_1, r_2 s_2 ... r_j s_j ... r_i s_i ... r_N s_N, t), \tag{19}$$

where $\theta = \frac{2\pi q}{\hbar c}\oiint \mathbf{M} \cdot \mathbf{dS} = \frac{2\pi q}{\hbar c}\Phi$, and $q$ is the charge.

Eq. (19) shows that the fractional statistics holds if and only if $0 < \theta < 2\pi$. However, if $n_\uparrow = n_\downarrow$ which could be attributed to the property of the AC phase, there would be no fractional statistics. Obviously, this is different from the situation of AB phase. The AC phase is very similar to AB phase in many respects, but different from the latter because of the non-Abelian nature of $SU(2)$ gauge structure. For example, the fractional statistics due to the AC phase is induced by the non-Abelian-Chern-Simons gauge field [10]. Although we cannot, as mentioned in the subsection 2.1, ensure that there exists purely $SU(2)$ gauge term in the Lagrangian (1), the Chern-Simons gauge term could be induced by the vacuum geometrical phase [33]. However, from previous discussion, we have confirmed that the fractional statistics could be achieved by the $U(1)$ Chern-Simons gauge field. Since $\mathbf{M} = \langle| \begin{pmatrix} \psi_\uparrow \\ \psi_\downarrow \end{pmatrix}^+ \boldsymbol{\sigma} \begin{pmatrix} \psi_\uparrow \\ \psi_\downarrow \end{pmatrix} |\rangle$ is constituted by up and down parts, the corresponding fields would be double Chern-Simons gauge fields. This is in agreement with the conclusion in Ref. [4].

## 6. Conclusion

In summary, three approaches, which refer to the Goldhaber-Anandan gauge theory (subsection 2.2), AC phase (section 3) and differential mapping (subsection 5.1) respectively, are introduced; use of them leads to a uniform expression for spin force. Especially, the approach of differential mapping removes zero spin Hall effect pointed out by Krachenko [13a], and prove that the C-D model is reasonable. The analysis to Goldhaber-Anandan gauge theory shows that our spin force is equivalent to results of Hirsch and Chudnovsky besides constant factors $3$ and $\frac{3}{2}$ respectively. The approach of AC phase implies that the spin force is a localized effect of AC phase; therefore, the spin Hall effect discussed by Hirsch [11a] and Chudnovsky [3a], in accordance with ours, is an intrinsic quantum mechanical property of spin. This is different from the approach of Berry phase in the study of intrinsic anomalous Hall effect, the latter performs more physical insight for universal anomalous Hall effect. However, the approach of AC phase uncovers that the AC phase is the most basic origin of spin Hall effect. The approach of differential mapping shows that neglect of the magnetic moment of nuclei, which originate from the fact that the mass of nuclei far exceed that of the electron, leads to that the residual magnetic moment of electron results in an "effect" magnetic field which could induce the fractional statistics and spin Hall effect.